\documentclass[preprint,showpacs,preprintnumbers,amsmath,amssymb]{revtex4}

\usepackage{graphicx}
\usepackage{dcolumn}
\usepackage{bm}

\begin{document}
%
\preprint{APS/123-QED}
%
\title{Electronic states and quantum transport\\ in double-wall carbon nanotubes}
\author{Seiji Uryu}
\email{uryu@stat.phys.titech.ac.jp}
\affiliation{Semiconductors Laboratory, RIKEN (Institute of Physical and Chemical Research)\\
2-1 Hirosawa, Wako, Saitama, Japan, 351-0198}
\affiliation{Condensed-Matter Theory Laboratory, RIKEN (Institute of Physical and Chemical Research)\\
2-1 Hirosawa, Wako, Saitama, Japan, 351-0198}
\date{\today}
%
\begin{abstract}
Electronic states and transport properties of double-wall carbon nanotubes without impurities are studied in a systematic manner.
It is revealed that scattering in the bulk is negligible and the number of channels determines the average conductance.
In the case of general incommensurate tubes, separation of degenerated energy levels due to intertube transfer is suppressed in the energy region higher than the Fermi energy but not in the energy region lower than that.
Accordingly, in the former case, there are few effects of intertube transfer on the conductance, while in the latter case, separation of degenerated energy levels leads to large reduction of the conductance.
It is also found that in some cases antiresonance with edge states in inner tubes causes an anomalous conductance quantization, $G\!=\!e^2/\pi\hbar$, near the Fermi energy.
\end{abstract}
%
\maketitle
%
\section{\label{sec:level 1}Introduction}
%
Carbon nanotubes (CN's) are cylindrical honeycomb lattices consisting of carbon atoms \cite{Iijima 1991a}.
They are often synthesized as complexes such as concentric multitube systems called multiwall carbon nanotubes (MWCN's) and bundles consisting of single-wall carbon nanotubes (SWCN's).
In complexes of CN's, the lattice structures of tubes are not considered to correlate with each other because of the weak interactions between the tubes.
However, electrons transfer from one tube to another with a low probability.
Although these complex CN's have often been used in experiments, the effects of the intertube transfer of electrons are not well understood.
In this paper we study double-wall carbon nanotubes (DWCN's), which are coaxial two-tube systems and the simplest MWCN, in order to clarify the effects of the intertube transfer on electronic states and transport properties of MWCN's.
\par
%
Two characteristics of the structure of DWCN's are known.
One is that, as mentioned above, the lattice structures of inner and outer tubes are not correlated with each other.
This means that DWCN's generally have no translational symmetry, i.e., one tube is incommensurate with the other one \cite{Kociak et al 2002a,Zuo et al 2003a}.
The other is that the difference between the radius of the inner tube and that of the outer one is about $\Delta R\!\approx\!3.6\AA$ independent of the circumference of the DWCN \cite{Bandow et al 2000a}.
Synthesis of DWCN's can be selectively performed \cite{Bacsa et al 2000a,Smith and Luzzi 2000a}.
However, few measurements of electrical transport in DWCN's have been carried out so far \cite{Kociak et al 2002a}.
\par
%
On the other hand, numerous experiments on transport properties of SWCN's, MWCN's, and carbon-nanotube bundles have been performed \cite{Dai et al 1996a,Ebbesen et al 1996a,Langer et al 1996a,Tans et al 1997a,Tsukagoshi et al 1999a,Kanda et al 2002a}.
A few experiments showed conductance quantization in SWCN's \cite{Kong et al 2001a} or MWCN's \cite{Frank et al 1998a,Urbina et al 2003a}, indicating the possibility of ballistic transport in the bulk of CN's and the realization of negligibly small contact resistance.
In one of the experiments using MWCN's, anomalous conductance quantization such as $G\!=\!e^2/\pi\hbar$ and $3e^2/\pi\hbar$ was reported\cite{Frank et al 1998a} but the origin is not yet well understood.
\par
%
A representative experimental setup is that in which source and drain electrodes are placed on or beneath CN's.
It has been reported that in the setup using MWCN's, electrodes are in contact with only one or two of the outermost tubes \cite{Bachtold et al 1999a}.
In another setup, one end of a CN is attached to the tip of a microscope and the other end is immersed into liquid metal \cite{Kociak et al 2002a,Frank et al 1998a}.
\par
%
Theoretically, energy bands for commensurate DWCN's and/or MWCN's have been calculated and energy-level separation due to the intertube transfer \cite{Saito et al 1993a,Kwon and Tomanek 1998a,Miyamoto et al 2001a,Kim et al 2001a} and resulting reductions of the conductance have been reported \cite{Sanvito et al 2000a,Kim et al 2001a}.
For some incommensurate DWCN's, the density of states was calculated and a weak effect of the intertube transfer near the Fermi energy has been reported \cite{Lambin et al 2000a}.
A study on level statistics in incommensurate DWCN's showed energy spacing distributions that depend on the energy windows \cite{Ahn et al 2003a}.
It was suggested that the intertube transfer in long incommensurate MWCN's becomes negligibly small \cite{Yoon et al 2002a} and the effect of the intertube transfer on the transport property is weak for an incommensurate DWCN and MWCN \cite{Roche et al 2001a}.
Although theoretical studies of MWCN's, including DWCN's, are proceeding as described above, overall features, particularly those on transport properties of incommensurate MWCN's, are not yet well understood.
\par
%
It is a characteristic property of graphite that for the Fermi energy, there exist localized states at the edges \cite{Fujita et al 1996a,Nakada et al 1996a}.
Therefore, states can also localize at open edges of CN's.
It is expected that in the cases of MWCN's these edge states of inner tubes affect current-carrying channels when the current in outer tubes passes near the edges.
\par
%
Although there is generally no translational symmetry in DWCN's, we can still use the concept of channels because the intertube transfer is sufficiently weak to be treated as a perturbation.
Then, it is considered that the electrical transport in DWCN's without impurities is governed by two factors.
One is the number of channels determined by energy bands in the absence of the intertube transfer and the separation of degenerated energy levels due to it.
The other is scattering from the intertube transfer due to incommensurability between the lattices of the two tubes.
The purpose of this paper is to clarify the overall electronic states and transport properties of MWCN's without impurities by studying DWCN's from this point of view.
\par
%
The paper is organized as follows.
The model and method are described in Sec.~\ref{sec:level 2}.
We show calculated results for electronic states in Sec.~\ref{sec:level 3} and those for transport properties in Sec.~\ref{sec:level 4}.
The results are discussed in Sec.~\ref{sec:level 5}.
The summary is given in Sec.~\ref{sec:level 6}.
\par
%
\begin{figure*}
\scalebox{0.55}{\includegraphics{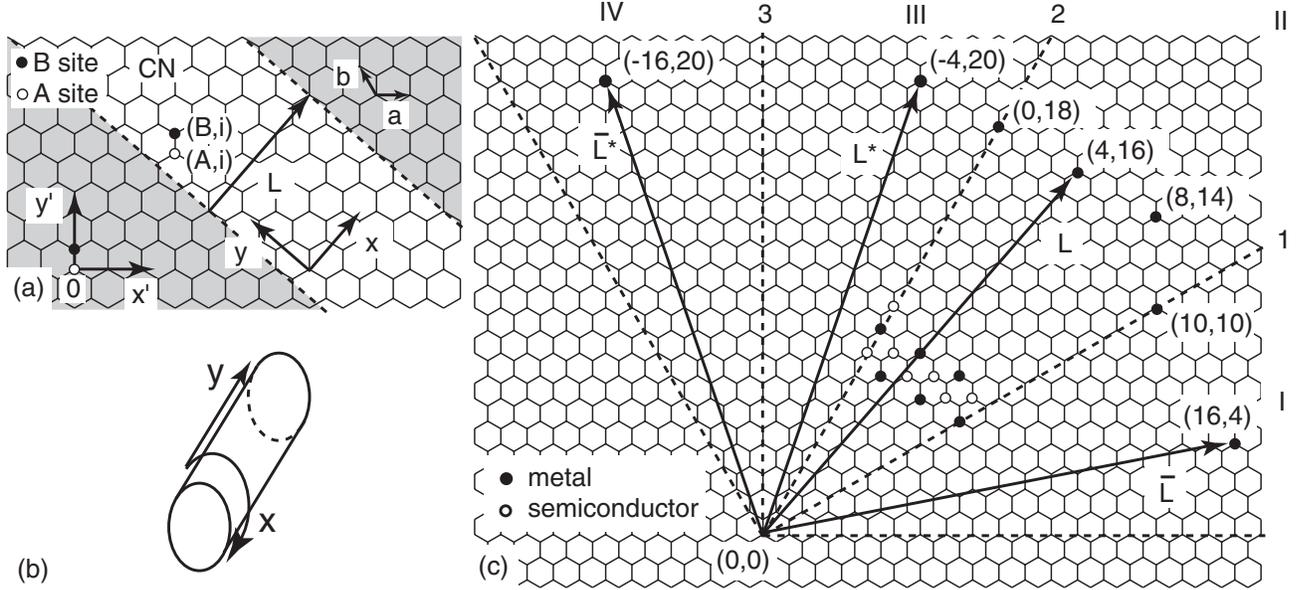}}
\caption{\label{fig:structures} 
Schematic of (a) and (c) 2D graphite sheets and (b) CN.
In (a) A and B sites in the $i$th unit cell of the graphite sheet are shown by (A,$i$) and (B,$i$), respectively.
The origin of the $x'y'$ coordinate system is chosen at an A site.
In (c) calculated CN's are shown by filled circles for metallic tubes and open circles for semiconducting tubes.
}
\end{figure*}
%
\section{\label{sec:level 2}Model and method}
%
A CN is considered as a two-dimensional (2D) graphite sheet with the periodic boundary condition for sites connected by a chiral vector ${\bf L}$ that determines the circumference.
Figure~\ref{fig:structures}(a) shows a schematic of the 2D graphite sheet.
A CN can be specified by chiral vector ${\bf L}\!=\!n_a{\bf a}\!+\!n_b{\bf b}$ with ${\bf a}\!=\!a(1,0)$ and ${\bf b}\!=\!a(-1/2,\sqrt{3}/2)$ being the lattice vectors, $a$ the lattice constant, and $n_a$ and $n_b$ integers.
\par
%
A CN is usually identified by a set of two integers defined as $(n_a\!-\!n_b,n_b)$.
When $(n_a\!-\!n_b)\!-\!n_b$ is an integer multiple of three, the tube is metallic; otherwise, the tube is semiconducting \cite{Mintmire et al 1992a,Hamada et al 1992a,Saito et al 1992a}.
Almost all tubes are chiral, however, only the $(n,n)$ tube and $(0,n)$ tube where $n$ is an integer are achiral.
The former is an armchair tube and the latter a zigzag tube.
\par
%
In the following, we denote the coordinate system on the 2D graphite sheet as $(x',y')$.
The A and B sites in the $i$th unit cell of the 2D graphite sheet are chosen as shown in Fig.~\ref{fig:structures}(a) and are denoted as (A,$i$) and (B,$i$), respectively, for text and A$i$ and B$i$, respectively, for subscripts.
For CN's we denote the circumference direction as the $x$ direction and the tube axis as the $y$ direction (see Figs.~\ref{fig:structures}(a) and (b)).
\par
%
All tubes can be specified by chiral vectors in a one-third area of the 2D graphite sheet as shown in Fig.~\ref{fig:structures}(c).
Let us equally divide this area into four regions with three straight boundaries 1, 2, and 3 passing through the origin and we name the four regions I, II, III, and IV as shown in the figure.
Boundaries 1 and 3 are the directions of the chiral vectors for armchair tubes and boundary 2 is that for zigzag tubes.
\par
%
Consider the four tubes specified by chiral vectors ${\bf L}$, ${\bf L}^*$, $\bar{\bf L}$, and $\bar{\bf L}^*$.
An example is shown in Fig.~\ref{fig:structures}(c).
The vectors ${\bf L}^*$ and $\bar{\bf L}^*$ are mirror symmetric to ${\bf L}$ and $\bar{\bf L}$, respectively, with respect to boundary 2, i.e., the direction for zigzag tubes.
The vector $\bar{\bf L}$ ($\bar{\bf L}^*$) is mirror symmetric to ${\bf L}$ (${\bf L}^*$) with respect to boundary 1 (3), i.e., the direction for armchair tubes.
The lattice structures of these tubes are mirror symmetric to each other.
In fact, the lattice structures of the tubes for ${\bf L}^*$ and $\bar{\bf L}^*$ are the same as those for ${\bf L}$ and $\bar{\bf L}$, respectively, with inversion of the circumference direction $x\!\rightarrow\!-x$ and the lattice structures of the tubes for $\bar{\bf L}$ and $\bar{\bf L}^*$ are the same as those for ${\bf L}$ and ${\bf L}^*$, respectively, with inversion of the axis direction $y\!\rightarrow\!-y$.
Therefore, the tubes for $\bar{\bf L}^*$ and $\bar{\bf L}$ are essentially the same as those for ${\bf L}$ and ${\bf L}^*$, respectively.
Once the properties of a tube are known, those of tubes mirror symmetric to it are also easily determined.
\par
%
In the case of DWCN's, however, the situation changes.
For a fixed inner tube, the lattice structure of the DWCN with the outer tube for $\bar{\bf L}^*$ ($\bar{\bf L}$) is no longer the same as that with the outer tube for ${\bf L}$ (${\bf L}^*$).
Therefore, we must survey tubes for all regions I to IV.
In order to systematically investigate all DWCN's with approximately the same circumference, we first take various combinations of inner and outer tubes in region II.
Then, fixing inner tubes to those in region II, we compare the results with those of other DWCN's with outer tubes in regions I, III, and IV that are mirror symmetric to or the same as one of the former outer tubes.
We will see in Sec.~\ref{sec:level 4-4} that the grouping of CN's into these four regions simplifies the investigation of DWCN's.
\par
%
We use the one-particle tight-binding model including the $\pi$ orbital in which no impurity potentials are considered.
The nearest-neighbor intratube transfers are taken into account, while the intertube transfers from one site to not only its nearest-neighbor sites but also other sites within hopping range are taken into account.
\par
%
The intertube transfer integral between the $(S_1,i)$ site in one tube and the $(S_2,j)$ site in the other tube, where $S_1,S_2\!=\!$ A or B,  is chosen as \cite{Nakanishi and Ando 2001a}
%
\begin{eqnarray}\label{eq:intertube interaction}
V_{S_1i,S_2j}\!&=&\!\alpha\gamma_1\exp\Bigl(-\frac{d-c/2}{\delta}\Bigr)\Bigl(\frac{{\bf p}_1\cdot{\bf d}}{d}\Bigr)\Bigl(\frac{{\bf p}_2\cdot{\bf d}}{d}\Bigr)\nonumber\\
&&-\gamma_0\exp\Bigl(-\frac{d-a_0}{\delta}\Bigr)
\Bigl[({\bf p}_1\cdot{\bf e})({\bf p}_2\cdot{\bf e})+({\bf p}_1\cdot{\bf f})({\bf p}_2\cdot{\bf f})\Bigr].
\end{eqnarray}
%
In Eq.(\ref{eq:intertube interaction}), $-\gamma_0$ and $\gamma_1$ are parameters of the effective model of graphite \cite{McClure 1957a,Slonczewski and Weiss 1958a} where $-\gamma_0$ is the transfer integral between nearest-neighbor sites for the same layer and $\gamma_1$ is that for neighboring layers.
Parameter $\alpha$ compensates the deviation of the transfer from $\gamma_1$ due to summation over sites in the hopping range.
Parameter $a_0/a\!=\!1/\sqrt{3}$ is the length of C-C bonds, $c/a\!=\!2.72$ the lattice constant along the $c$ axis of graphite, and $\delta$ the decay rate of the $\pi$ orbital.
Vectors ${\bf p}_1$ and ${\bf p}_2$ are the unit vectors directed to the $\pi$ orbital at $(S_1,i)$ and $(S_2,j)$ sites, respectively, ${\bf d}$ the vector connecting the two sites, and ${\bf e}$ and ${\bf f}$ the unit vectors perpendicular to ${\bf d}$ and to each other.
\par
\begin{figure}
\scalebox{0.45}{\includegraphics{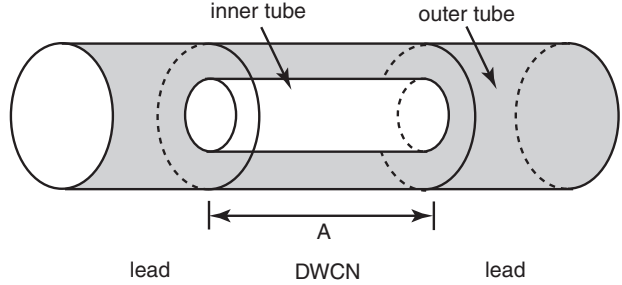}}
\caption{\label{fig:two-terminal system} Schematic of a two-terminal DWCN system.
The inner tube is finite with length $A$ and open edges and the outer tube is adequately long and connected to reservoirs and plays the role of ideal leads.}
\end{figure}

%
In calculations of transport coefficients, we use the two-terminal system shown in Fig.~\ref{fig:two-terminal system}.
The inner tube is finite with length $A$ and armchair and/or zigzag open edges.
The adequately long outer tube is connected to reservoirs and plays the role of ideal leads in the region without the inner tube.
This corresponds to the case where electrodes are attached to only outer tubes.
Scattering ($S$) matrices are calculated by the recursive Green's function method\cite{Ando 1991a}, and we calculate the conductance using the Landauer-B\"uttiker formula\cite{Landauer 1957a,Buttiker et al 1985a}.
\par
%
%
In the following calculations, we use the parameters in Eq.(\ref{eq:intertube interaction}): $\gamma_1/\gamma_0\!=\!0.119$ \cite{Nagayoshi et al 1976a}, $\delta/a\!=\!0.185$ \cite{Lambin et al 2000a}, and $\alpha\!=\!1.4$.
The value of $\alpha$ is chosen by fitting the energy dispersion of graphite in the $c$-axis direction calculated from Eq.(\ref{eq:intertube interaction}) to those in the effective model \cite{McClure 1957a,Slonczewski and Weiss 1958a}.
\par
%
Numerical calculations are performed for the following DWCN's.
Among CN's in region II in Fig.~\ref{fig:structures}(c), metallic (4,16), (0,18), (8,14), and (10,10) tubes are chosen as outer tubes and metallic (3,6), (1,7), (5,5), (0,9), (2,8), and (4,7) tubes and semiconducting (0,8), (0,10), (1,8), (2,7), (3,7), (4,6), and (5,6) tubes are chosen as inner tubes.
Then, the radii of the outer tubes are about $2.9a$ and those of the inner tubes about $1.4a$.
Among all possible combinations of the above tubes, we perform calculations for DWCN's in which the difference between the two radii satisfies $1.3\!\alt\!\Delta R/a\!\alt\!1.6$, in line with the experiment \cite{Bandow et al 2000a}.
Calculated results for these DWCN's are given in Secs.~\ref{sec:level 3} and ~\ref{sec:level 4-1} to \ref{sec:level 4-3}.
\par
%
We perform calculations for three more outer tubes, (16,4), ($-$4,20), and ($-$16,20) tubes, that correspond to tubes of chiral vectors $\bar{\bf L}$, ${\bf L}^*$, and $\bar{\bf L}^*$, respectively, where the chiral vector of the (4,16) tube is chosen as ${\bf L}$, as shown in Fig.~\ref{fig:structures}(c).
These calculations are presented in Sec.~\ref{sec:level 4-4}.
\par
%
\section{\label{sec:level 3}Electronic states}
%
In this section, we study the electronic states of incommensurate DWCN's.
The lowest order effect of the intertube transfer on the energy levels is the separation of degenerated energy levels that leads to a reduction of the number of channels in the DWCN region.
\par
%
First, consider the two 2D graphite sheets stacked incommensurately.
We call them layers 1 and 2.
The two sheets are parallel to the $x'y'$ plane and perpendicular to the $z'$ axis.
Without loss of generality, we can obtain two incommensurately stacked sheets in the following way.
First, prepare two commensurately stacked sheets in which the lattice points of layer 1 are located at the same positions as those of layer 2 with respect to the $x'y'$ coordinate, but the $z'$ coordinate is different.
The origin is chosen at an A site of layer 1, i.e., the (A,0) site.
The $x'y'$ coordinate system, which is used in the following, is defined  as shown in Fig.~\ref{fig:structures}(a) at this stage.
Next, rotate layer 2 around the $z'$ axis by an angle $\theta$.
We shall calculate the separation width of degenerated energy levels in this system.
\par
%
Because the interlayer transfer is a weak effect and can be treated as a perturbation, states are specified by eigenstates of a single 2D graphite sheet.
In the nearest-neighbor tight-binding model, the Schr\"odinger equation for the 2D graphite sheet is
%
\begin{eqnarray}\label{eq:Schrodinger eq. of graphite A}
-\gamma_0\Bigl[1\!+\!e^{i(-k_{x'}-\sqrt{3}k_{y'})a/2}\!+\!e^{i(k_{x'}-\sqrt{3}k_{y'})a/2}\Bigr]C_{Bi}&\!=\!&EC_{Ai},\\
-\gamma_0\Bigl[1\!+\!e^{i(-k_{x'}+\sqrt{3}k_{y'})a/2}\!+\!e^{i(k_{x'}+\sqrt{3}k_{y'})a/2}\Bigr]C_{Ai}&\!=\!&EC_{Bi},
\label{eq:Schrodinger eq. of graphite B}
\end{eqnarray}
%
where we use ${\bf R}_{Bi}\!-\!{\bf R}_{Ai}\!=\!{\bf R}_{B0}\!=\!(0,a/\sqrt{3})$ with ${\bf R}_{Ai}$ and ${\bf R}_{Bi}$ being the coordinates of (A,$i$) and (B,$i$) sites, respectively, (see Fig.~\ref{fig:structures}(a)), $C_{Ai}$ and $C_{Bi}$ are the wave functions, and $E$ is the eigenenergy. 
The Bloch theorem is used with the wave vector ${\bf k}$.
The Fermi energy of intrinsic CN's is chosen as the energy origin throughout this paper.
The energy dispersion is given by
%
\begin{eqnarray}\label{eq:energy dispersion}
&&E_\pm\!=\!\pm\gamma_0\Bigl[1\!+\!4\cos\Bigl(\frac{k_{x'}a}{2}\Bigr)\cos\Bigl(\frac{\sqrt{3}k_{y'}a}{2}\Bigr)\!+\!4\cos^2\Bigl(\frac{k_{x'}a}{2}\Bigr)\Bigr]^{1/2}\!\!\!\!\!.
\end{eqnarray}
%
Energy bands consist of the conduction band for $0\!\le\!E/\gamma_0\!\le\!3$ and the valence band for $-3\!\le\!E/\gamma_0\!\le\!0$.
The conduction and valence bands are the maximum and minimum, respectively, at ${\bf k}\!=\!0$.
They are mirror symmetric to each other with respect to $E\!=\!0$ and have six-fold rotational symmetry about ${\bf k}\!=\!0$.
The wave functions are given by
%
\begin{eqnarray}\label{eq:wave function A}
&&C_{Ai}=C_{A0}e^{i{\bf k}\cdot{\bf R}_{Ai}},\\
&&C_{Bi}=C_{B0}e^{i{\bf k}\cdot{\bf R}_{Ai}}.\label{eq:wave function B}
\end{eqnarray}
%
The wave functions at (A,0) and (B,0) sites are related to each other by the following:
%
\begin{equation}\label{eq:BA site relation}
C_{B0}=-\frac{\gamma_0}{E}\Bigl[
1+2\cos\Bigl(\frac{k_{x'}a}{2}\Bigr)e^{i\sqrt{3}k_{y'}a/2}\Bigr]
C_{A0}.
\end{equation}
%
Therefore, for $ka\ll 1$, i.e., $|E|/\gamma_0\!\sim\!3$, we have
%
\begin{equation}
C_{B0}\approx-{\rm sgn}(E)C_{A0}e^{i{\bf k}\cdot{\bf R}_{B0}},
\label{eq:B site wave function for ka<<1}
\end{equation}
%
where sgn($E$) denotes the sign of $E$.
\par
%
Next, consider the matrix element of the interlayer transfer.
Using Eqs. (\ref{eq:wave function A}) and (\ref{eq:wave function B}) the matrix element of transition from a state with ${\bar{\bf k}}_2$ at energy $E_2$ in layer 2 to a state with ${\bf k}_1$ at energy $E_1$ in layer 1 is given by
%
\begin{eqnarray}\label{eq:transition amplitude 1}
\langle {\bf k}_1|{\hat V}|{\bar{\bf k}}_2\rangle&=&\sum_{S_2j}\sum_{S_1i}C_{S_1i}^*V_{S_1i,S_2j}D_{S_2j}\nonumber\\
&=&\sum_j\sum_i(C_{A0}^*V_{Ai,Aj}+C_{B0}^*V_{Bi,Aj})\nonumber\\
&&\times e^{-i{\bf k_1}\cdot({\bf R}_{Ai}-{\bar{\bf R}}_{Aj})}D_{A0}e^{i({\bar{\bf k}}_2-{\bf k}_1)\cdot{\bar{\bf R}}_{Aj}}\nonumber\\
&&+\sum_j\sum_i(C_{A0}^*V_{Ai,Bj}+C_{B0}^*V_{Bi,Bj})\nonumber\\
&&\times e^{-i{\bf k}_1\cdot({\bf R}_{Ai}-{\bar{\bf R}}_{Bj})}D_{B0}e^{i({\bar{\bf k}}_2-{\bf k}_1)\cdot{\bar{\bf R}}_{Bj}-i{\bar{\bf k}}_2\cdot{\bar{\bf R}}_{B0}},
\end{eqnarray}
%
where ${\hat V}$ is the operator of the interlayer transfer, $C_{S_1i}$ ($D_{S_2j}$) denotes the wave function of layer 1 (2) at the $(S_1,i)$ site ($(S_2,j)$ site), and ${\bf R}$ and ${\bar{\bf R}}$ are positions of sites in layers 1 and 2, respectively.
\par
%
Let us consider the case of $k_1a\!\ll\!1$ and $k_2a\!\ll\!1$, i.e., $|E_1|/\gamma_0\!\sim\!|E_2|/\gamma_0\!\sim\!3$.
Then, the range of the interlayer transfer is much smaller than the wavelength.
Therefore, the phase factors $\exp[-i{\bf k_1}\cdot({\bf R}_{Ai}-{\bar{\bf R}}_{Aj})]$ and $\exp[-i{\bf k}_1\cdot({\bf R}_{Ai}-{\bar{\bf R}}_{Bj})]$ in Eq. (\ref{eq:transition amplitude 1}) can be set to zero.
Then, the effective interlayer transfer integral from the $(S_2,j)$ site of layer 2 to $S_1$ sites of layer 1 can be considered to be $\sum_iV_{S_1i,S_2j}$.
Similarly, that from the $(S_1,i)$ site of layer 1 to $S_2$ sites of layer 2 can be considered to be $\sum_jV^*_{S_1i,S_2j}$.
\par
%
\begin{figure}
\scalebox{0.45}{\includegraphics{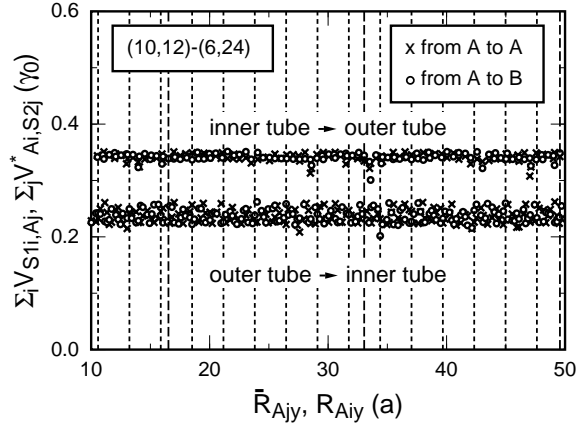}}
\caption{\label{fig:position dependence of intertube transfer} Position dependence of the effective intertube transfer integral in the case of (10,12)-(6,24) DWCN.
Horizontal axis is the $y$ coordinate of sites from which electrons transfer to the other tube.
Results for the transfers from A sites of one tube to A sites (open circles) and B sites (crosses) of the other tube are plotted.
The results for sites whose azimuthal angles are between 0 and $\pi/3$ are shown for simplicity.
}
\end{figure}
%
Figure~\ref{fig:position dependence of intertube transfer} shows $\sum_iV_{S_1i,Aj}$ for (10,12)-(6,24) DWCN which is the effective transfer integral from the (A,$j$) site of the inner (10,12) tube to $S_1$ sites of the outer (6,24) tube.
The horizontal axis is the $y$ coordinate ${\bar R}_{Ajy}$ of the (A,$j$) site of the inner tube, where the $y$ direction is parallel to the tube axis.
The effective transfer integral from the (A,$i$) site of the outer tube to $S_2$ sites of the inner tube, $\sum_jV^*_{Ai,S_2j}$, is also similarly plotted.
The results for the transfer from B sites are similar to those for the transfer from A sites and are not shown in the figure.
The intervals between dotted lines indicate the length of the unit cell for the outer tube and those between dashed lines that for the inner tube.
\par
%
The result shows that the effective intertube transfer integral negligibly depends on the position and is considered approximately constant.
This is because electrons can transfer to not only the nearest-neighbor site but also other sites in the hopping range.
Therefore, the intertube transfer can be considered to conserve the momentum.
There is no periodicity, as expected.
The effective transfer integral from the outer tube to the inner tube is smaller and fluctuates more than those from the inner tube to the outer tube.
This is because the diameter of the inner tube is smaller than that of the outer tube.
When the circumference of the DWCN is increased, the difference of the effective transfer integral between the two cases becomes smaller.
\par
%
Therefore, in the case of $k_1a\!\ll\!1$ and $k_2a\!\ll\!1$, important matrix elements in Eq. (\ref{eq:transition amplitude 1}) are those for ${\bf k}_1\!=\!{\bar{\bf k}}_2\!=\!{\bf k}$, i.e.,
%
\begin{eqnarray}\label{eq:transition amplitude 2}
\langle {\bf k}|{\hat V}|{\bf k}\rangle&\approx&\sum_{ij}\bigl[V_{Ai,Aj}-{\rm sgn}(E_1)V_{Bi,Aj}\nonumber\\
&&-{\rm sgn}(E_2)V_{Ai,Bj}+{\rm sgn}(E_1E_2)V_{Bi,Bj}\bigr]C_{A0}^*D_{A0},
\end{eqnarray}
%
where Eq.~(\ref{eq:B site wave function for ka<<1}) is used.
It should be noted that the eigenenergy $E_1$ of state with ${\bf k}$ in layer 1 is not generally equal to that in layer 2, $E_2$, because the lattice of layer 2 is rotated.
Since the lattice of layer 1 does not correlate with that of layer 2, we may make the following assumption for incommensurate cases:
%
\begin{equation}\label{eq:assumption of randomness}
\sum_{ij}V_{Ai,Aj}=\sum_{ij}V_{Bi,Aj}=\sum_{ij}V_{Ai,Bj}=\sum_{ij}V_{Bi,Bj}.
\end{equation}
%
Finally the matrix element is given as
%
\begin{eqnarray}\label{eq:matrix element}
\langle {\bf k}|{\hat V}|{\bf k}\rangle \approx \left\{
\begin{array}{cc}
4\sum_{ij}V_{Ai,Aj}C_{A0}^*D_{A0} & \quad\quad {\rm for}\quad E_1<0, E_2<0\\
0 & \quad\quad {\rm otherwise}
\end{array}
\right..
\end{eqnarray}
%
\par
%
We are particularly interested in the separation width of degenerated energy levels that is approximately given by $2|\langle {\bf k}|{\hat V}|{\bf k}\rangle|$ for ${\bf k}$'s satisfying $E_1\!=\!E_2\!=\!E$.
From the energy dispersion relation Eq.(\ref{eq:energy dispersion}), this is realized for ${\bf k}$'s given by
%
\begin{equation}\label{eq:degenerate point 1}
(k_{x'},k_{y'})=\Bigl(k\cos\Bigl(\frac{\theta}{2}+\frac{n\pi}{3}\Bigr),k\sin\Bigl(\frac{\theta}{2}+\frac{n\pi}{3}\Bigr)\Bigr),
\end{equation}
%
or
%
\begin{equation}\label{eq:degenerate point 2}
(k_{x'},k_{y'})=\Bigl(-k\sin\Bigl(\frac{\theta}{2}+\frac{n\pi}{3}\Bigr),k\cos\Bigl(\frac{\theta}{2}+\frac{n\pi}{3}\Bigr)\Bigr),
\end{equation}
%
where $n$ is an integer.
\par
%
From Eq.(\ref{eq:matrix element}), at degenerated energy levels in the conduction band where $E\!>\!0$, the matrix element vanishes and separation of degenerated energy levels is negligible.
In cases for the valence band where $E\!<\!0$, the separation width is given by twice the first line of Eq.(\ref{eq:matrix element}).
In spite of symmetric energy bands of each graphite sheet with respect to $E\!=\!0$, eigenenergy levels of incommensurately stacked graphite sheets are not symmetric due to the interlayer transfer which destroys the symmetry of the bipartite lattice.
\par
%
As ${\bf k}$ goes away from the case of $ka\!\ll\!1$, the matrix element increases in the conduction band and decreases in the valence band, because the phase factors $\exp[-i{\bf k_1}\cdot({\bf R}_{Ai}-{\bar{\bf R}}_{Aj})]$ and $\exp[-i{\bf k}_1\cdot({\bf R}_{Ai}-{\bar{\bf R}}_{Bj})]$ in Eq. (\ref{eq:transition amplitude 1}) break the cancellation in the conduction band and reduce the sum of the intertube transfer integral over sites in each hopping range in the valence band.
\par
%
\begin{figure}
\scalebox{0.45}{\includegraphics{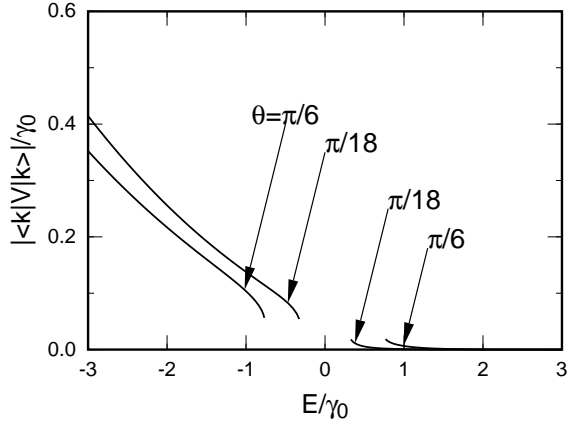}}
\caption{\label{fig:energy gap 2D incomm}
Absolute value of the interlayer transfer matrix element for two incommensurate graphite sheets for $\theta\!=\!\pi/6$ and $\pi/18$.
The distance between two layers is $c/2\!=\!1.36a$.
}
\end{figure}
%
\begin{figure}
\scalebox{0.45}{\includegraphics{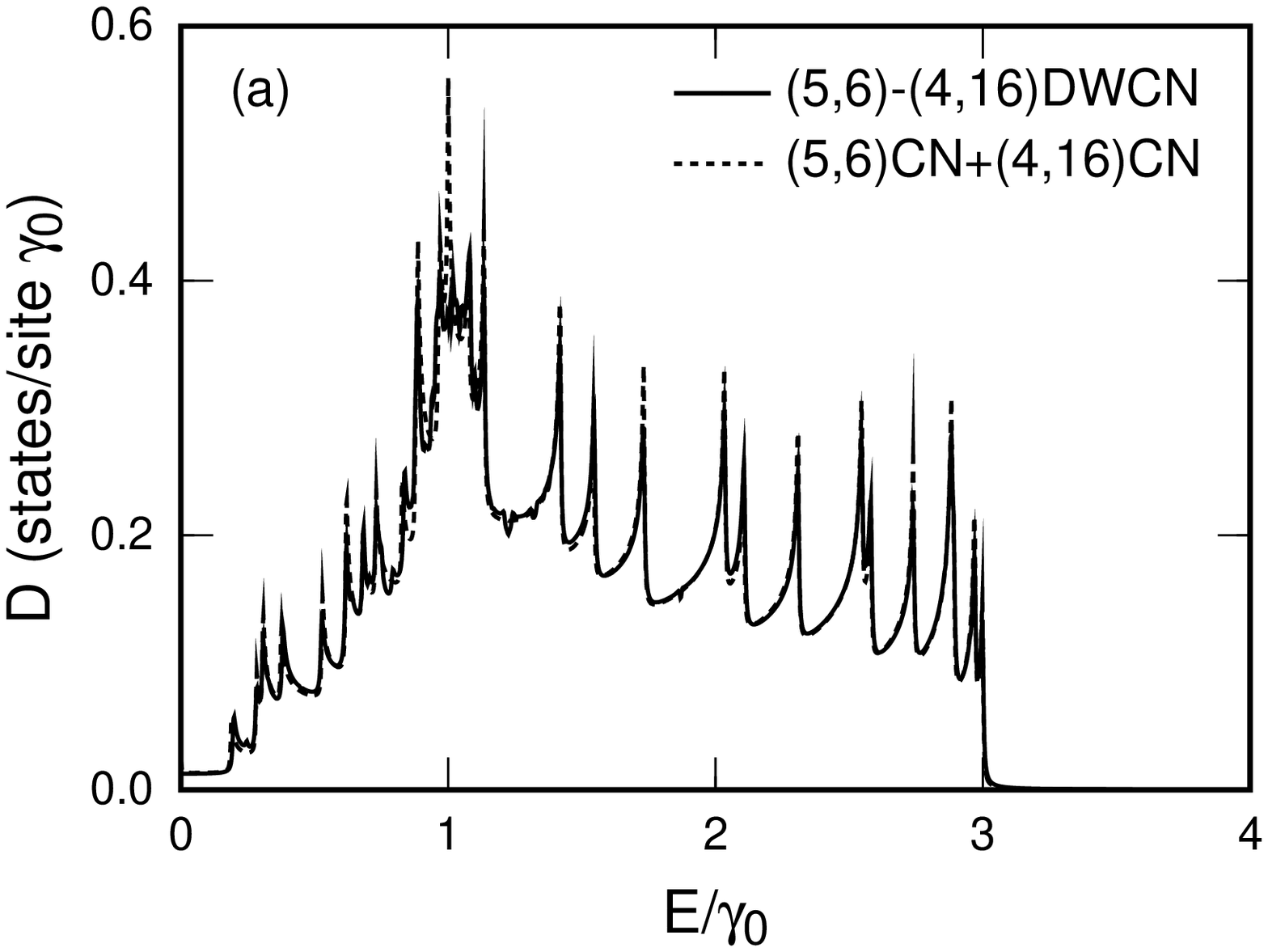}}
\scalebox{0.45}{\includegraphics{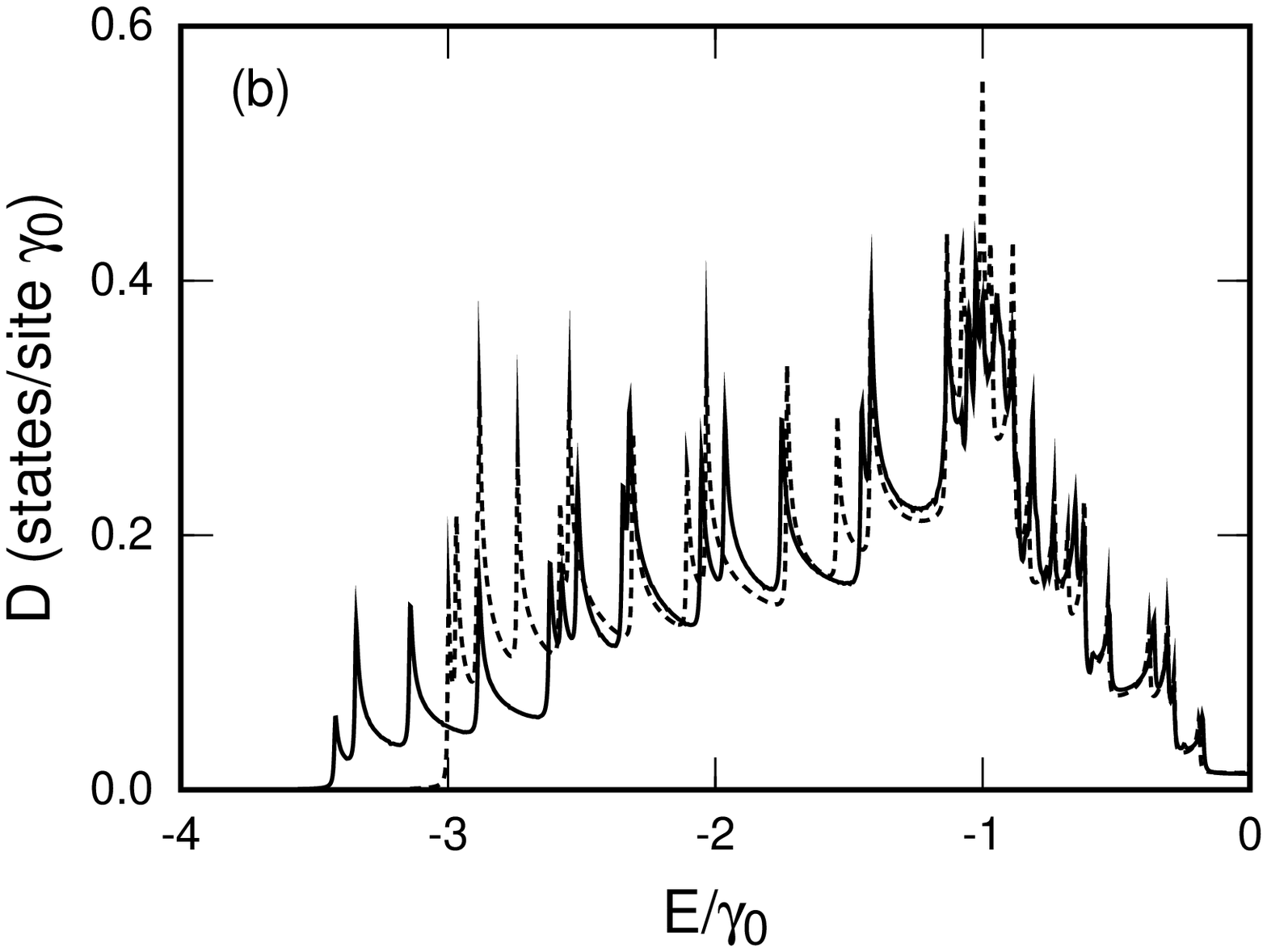}}
\caption{\label{fig:density of states DWCN incomm} 
Density of states of (5,6)-(4,16) DWCN (solid line) and sum of the density of states for (5,6) tube and that for (4,16) tube (dotted line) for (a) $E\!>\!0$ and (b) $E\!<\!0$.
}
\end{figure}
%
Figure~\ref{fig:energy gap 2D incomm} shows matrix element $|\langle {\bf k}|{\hat V}|{\bf k}\rangle|$ calculated using Eq.~(\ref{eq:transition amplitude 1}) and the interlayer transfer integral Eq.~(\ref{eq:intertube interaction}) for ${\bf k}$'s given by Eq.(\ref{eq:degenerate point 1}).
In the conduction band $E\!>\!0$, the matrix element is negligible not only for $E/\gamma_0\!\sim\!3$ but also for other energy ranges.
This indicates the validity of Eq.(\ref{eq:assumption of randomness}).
In the valence band $E\!<\!0$, the matrix element is maximum at $E/\gamma_0\!=\!-3$ and decreases with increasing energy.
\par
%
This result can be applied to incommensurate DWCN's.
As an example, the density of states $D$ for the (5,6)-(4,16) DWCN is shown by solid lines in Fig.~\ref{fig:density of states DWCN incomm}(a) for $E\!>\!0$ and (b) for $E\!<\!0$.
The sum of the density of states for (5,6) tube and that for (4,16) tube is also shown as dotted lines for reference.
For $E\!>\!0$, the density of states for the DWCN approximately follows the dotted line.
For $E\!<\!0$, the density of states for the DWCN deviates from the dotted line and additional van Hove singularities due to the separation of degenerated energy levels appear for $-1\!\alt\!E/\gamma_0\!\alt\!0$.
The deviation from the dotted line becomes greater with decreasing energy, and the bottom is pushed down below $E/\gamma_0\!=\!-3$.
These features are in good agreement with the results for two incommensurately stacked 2D graphite sheets, as expected.
\par
%
\begin{figure}
\scalebox{0.45}{\includegraphics{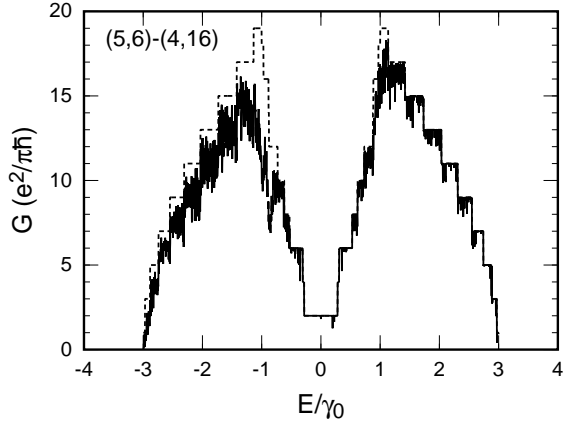}}
\caption{\label{fig:G of E (5,6)-(4,16) wide}
Energy dependence of the conductance for (5,6)-(4,16) DWCN.
The length is taken to be $A/a\!=\!809$, i.e., $A\!=\!200$nm.
Solid line shows the conductance of the DWCN and dotted line that of outer (4,16) tube.
}
\end{figure}
%
\begin{figure}
\scalebox{0.45}{\includegraphics{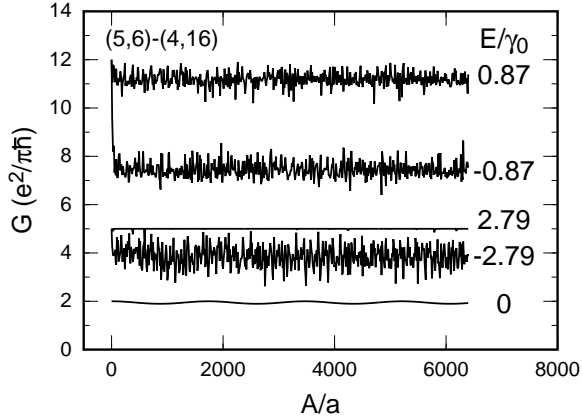}}
\caption{\label{fig:G of L (5,6)-(4,16)}
Length dependence of the conductance for (5,6)-(4,16) DWCN.
The numbers of channels in the outer tube are 2, 12, and 5 for $EL/2\pi\gamma\!=\!0$, $\pm0.87$, and $\pm2.79$, respectively.
}
\end{figure}
%
\section{\label{sec:level 4}Transport properties}
%
\subsection{\label{sec:level 4-1}Incommensurate DWCN's}
%
In this subsection, the calculated conductance of incommensurate DWCN's is presented.
As a typical result, Fig.~\ref{fig:G of E (5,6)-(4,16) wide} shows the energy dependence of the conductance for (5,6)-(4,16) DWCN.
The solid line is the conductance of the DWCN and the dotted line the conductance in the absence of the intertube transfer, that is, that of the outer tube.
In the case of $E\!<\!0$, the conductance of the DWCN is reduced as compared to that of the outer tube due to the effects of the intertube transfer.
On the other hand, for $E\!>\!0$ the conductance of the DWCN is very close to that of the outer tube, suggesting that the effect of the intertube transfer is weak.
These features hold not only for $|E|/\gamma_0\!\sim\!3$ but also for other energy regions.
Some deviations between the solid and dotted lines are seen at $E/\gamma_0\!\sim\!1$ but they are much smaller compared to those at $E/\gamma_0\!\sim\!-1$, the energy symmetric to $E/\gamma_0\!\sim\!1$.
\par
%
This is in good agreement with the result in the  previous section that separation of degenerated energy levels is suppressed for $E\!>\!0$ but not for $E\!<\!0$, because the separation leads to disappearance of the channels, and electrons coming through the corresponding channels in a lead are completely reflected at the interface between the lead and the DWCN region.
\par
%
The length dependence of the conductance clearly shows the above scattering mechanism, and that for the same (5,6)-(4,16) DWCN at $E/\gamma_0\!=\!0$, $\pm0.87$, and $\pm2.79$ is presented in Fig.~\ref{fig:G of L (5,6)-(4,16)}.
At $E/\gamma_0\!=\!\pm2.79$, i.e., $|E|/\gamma_0\!\sim\!3$, there are five channels in the outer tube.
The conductance for $E/\gamma_0\!=\!-2.79$ rapidly falls from $5e^2/\pi\hbar$ to about $4e^2/\pi\hbar$ near $A\!=\!0$ and weakly fluctuates around $4e^2/\pi\hbar$ thereafter.
This is the behavior expected for the separation of degenerated energy levels.
For $E/\gamma_0\!=\!2.79$, the conductance is about $5e^2/\pi\hbar$.
This is consistent with the suppression of separation of degenerated energy levels for $E\!>\!0$.
\par
%
Also in the case of $E/\gamma_0\!=\!\pm0.87$, which is far from $|E|/\gamma_0\!=\!3$, the result is similar.
There are twelve channels in the outer tube in this case.
The conductance at $E/\gamma_0\!=\!-0.87$ rapidly falls from $12e^2/\pi\hbar$ to about $7e^2/\pi\hbar$ in a short-length regime and shows small fluctuation around it thereafter.
On the other hand, the conductance at $E/\gamma_0\!=\!0.87$ shows a rapid and slight fall from $12e^2/\pi\hbar$ to about $11e^2/\pi\hbar$ and fluctuates weakly there.
This one-channel reduction at $E/\gamma_0\!=\!0.87\!>\!0$ is considered to arise from the fact that the energy is far from $E/\gamma_0\!=\!3$, but it is much smaller than the five-channel reduction at $E/\gamma_0\!=\!-0.87$.
The conductance at $E\!=\!0$ is about $2e^2/\pi\hbar$ and almost independent of the length due to the absence of degenerated energy levels.
\par
%
Apart from $A\!\sim\!0$, the conductance exhibits almost no length dependence except for small fluctuations.
Therefore, it is considered that scattering from the intertube transfer in bulk is weak.
This is consistent with the result in Fig.~\ref{fig:position dependence of intertube transfer} and suggests that the momentum conservation is satisfied not only for $|E|/\gamma_0\!\sim\!3$ but also for the other region.
For the momentum conservation in the intertube transfer prohibits backscattering from a channel of one tube to another of the same tube in bulk and only the current carrying through the outer tubes is measured in our case.
Thus, it can be said that average conductance is determined by the number of channels, that is, scattering at the interfaces between leads and DWCN regions.
\par
%
\begin{figure*}
\scalebox{0.31}{\includegraphics{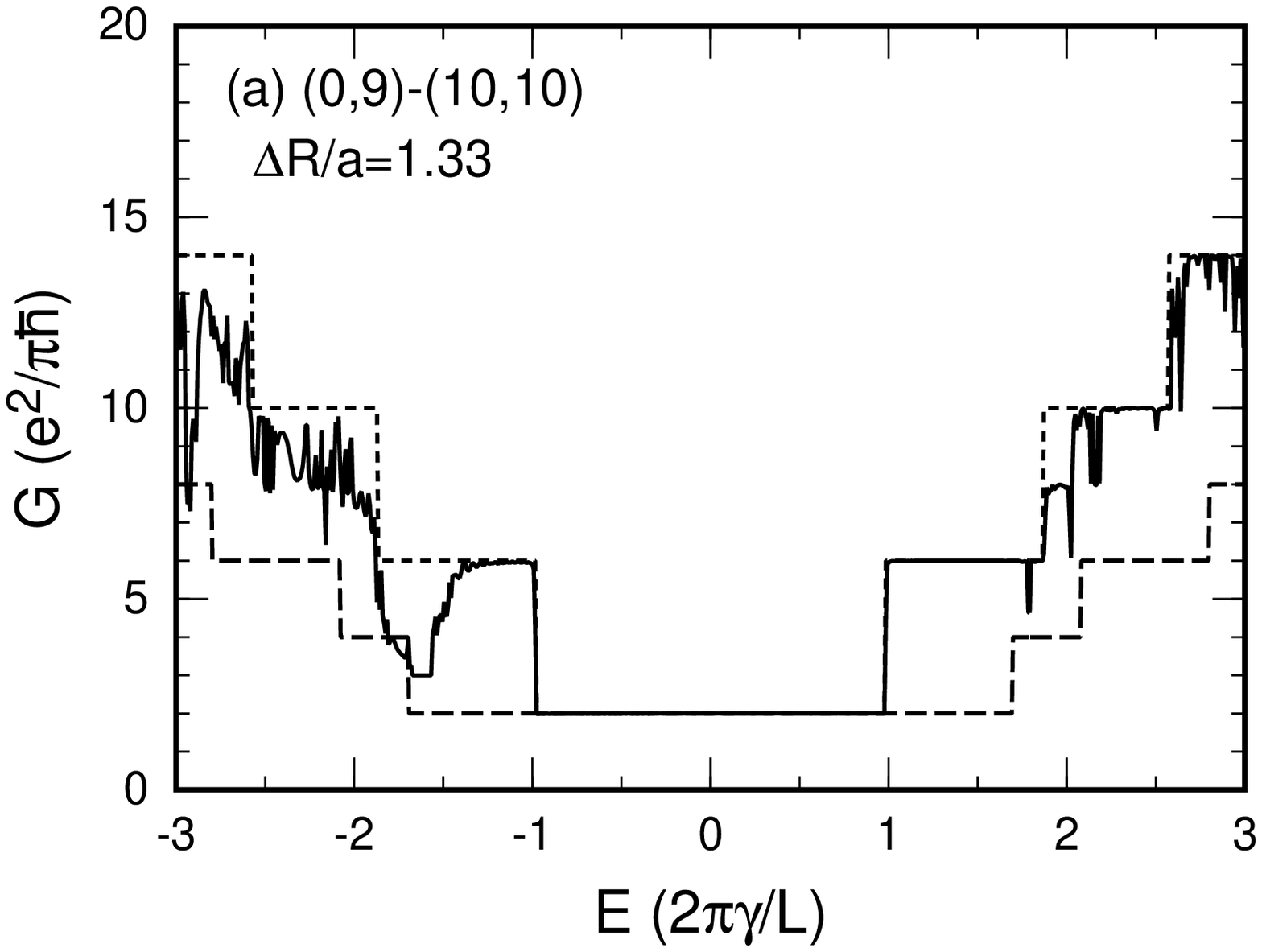}}
\scalebox{0.31}{\includegraphics{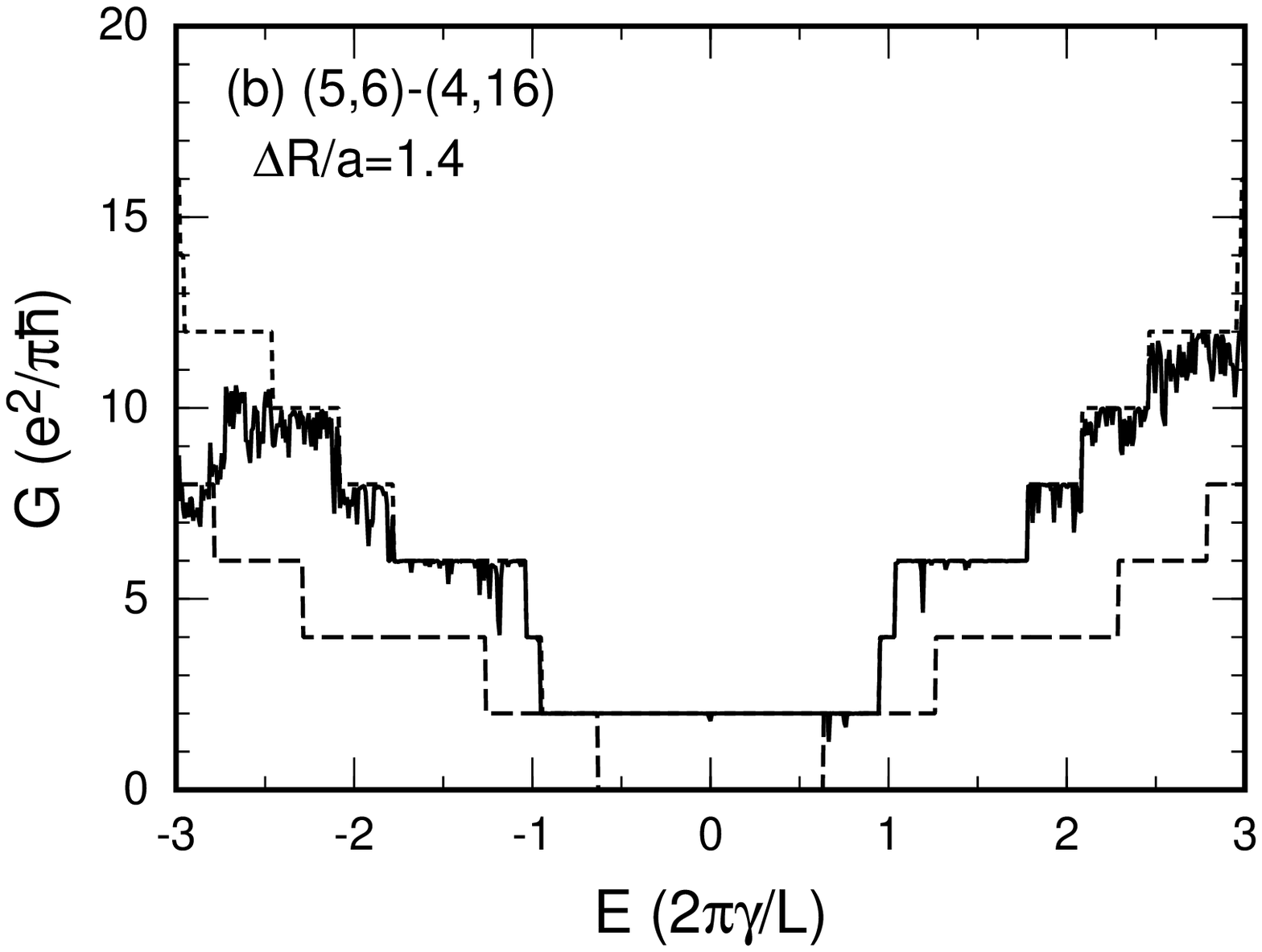}}
\scalebox{0.31}{\includegraphics{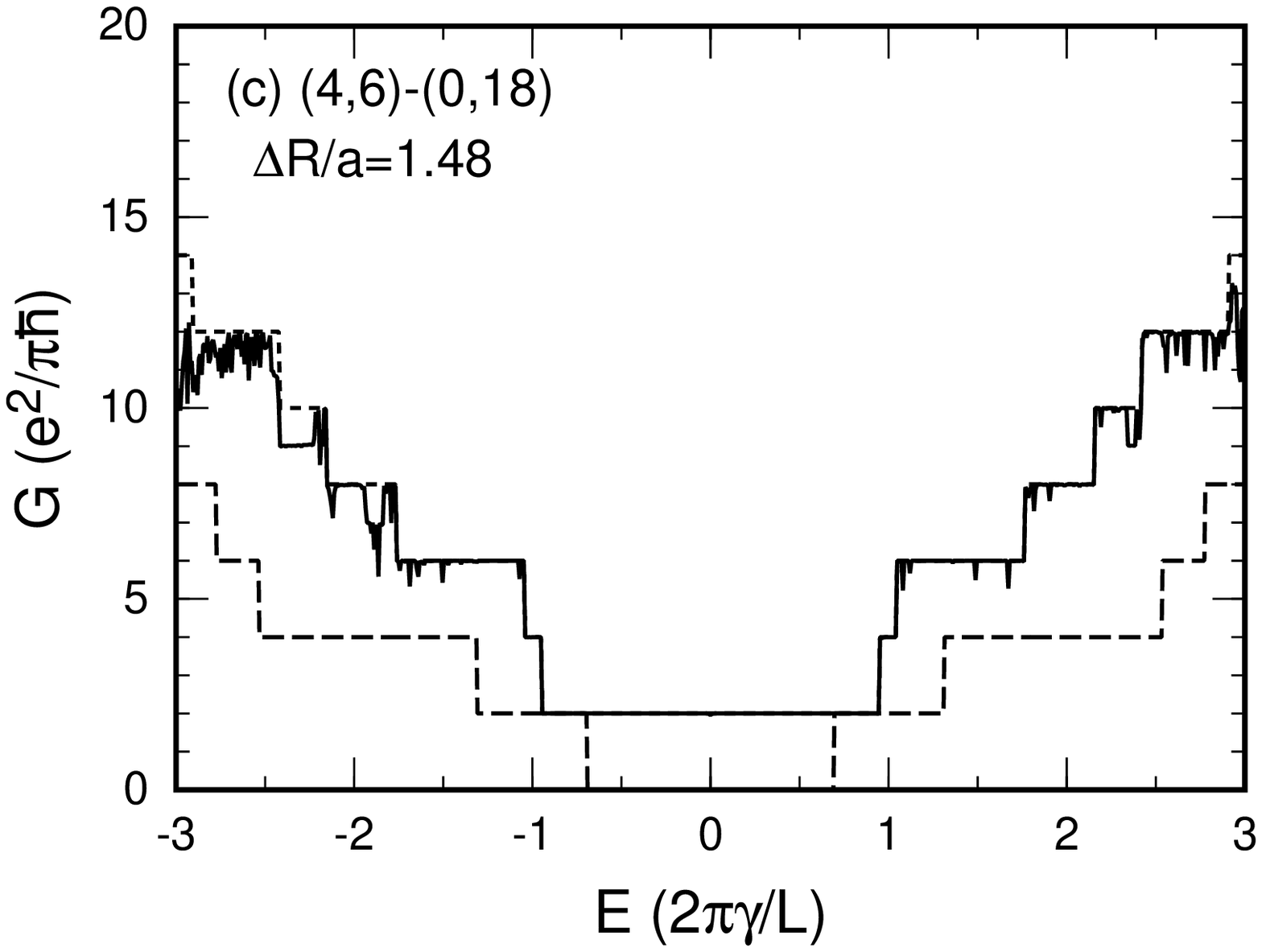}}
\caption{\label{fig:G of E incomm} 
Conductance of incommensurate DWCN's as a function of energy.
The length is taken as $A/a\!\sim\!800$, i.e., $A\!\sim\!200$nm.
Dotted lines indicate the number of channels of the outer tubes times $e^2/\pi\hbar$ and dashed lines those of the inner tubes times $e^2/\pi\hbar$.
In (a), the inner tube is metallic and in (c) and (b) semiconducting.
Among (a), (b), and (c), $\Delta R$ is smallest in (a) and largest in (c).
}
\end{figure*}
%
Figure~\ref{fig:G of E incomm} shows examples of the energy dependence of the conductance for various DWCN's around the Fermi energy.
Energy is measured in units of the bottom of the first excited subband $2\pi\gamma/L$ with $L$ being the circumference of the outer tube and $\gamma\!=\!\sqrt{3}\gamma_0a/2$.
Figure~\ref{fig:G of E incomm}(b) is an enlargement of Fig.~\ref{fig:G of E (5,6)-(4,16) wide} in the range of $-0.9\!\alt\!E/\gamma_0\!\alt\!0.9$.
The following common features are revealed.
First, large reductions of the conductance occur for $E\!<\!0$, rarely for $E\!>\!0$ and not in the energy region with two channels, $-1\!\alt\!EL/2\pi\gamma\!\alt\!1$.
Secondly, in the energy region with two channels, the transmission is nearly perfect, i.e., $G\!\approx\!2e^2/\pi\hbar$, and as the number of channels is increased, the conductance begins to fluctuate.
Thirdly, scattering becomes stronger as the difference between radii of inner and outer tubes $\Delta R$ becomes smaller.
\par
%
The first feature is due to the previous result that separation of degenerated levels is suppressed in $E\!>\!0$ but not in $E\!<\!0$ and to that degeneracy of energy levels rarely occurs at energy close to the Fermi energy.
The third feature is attributed to the simple fact that as $\Delta R$ becomes smaller, the intertube transfer integral becomes larger, as shown in Eq.(\ref{eq:intertube interaction}).
\par
%
The second feature is the conductance fluctuation.
It is possible that weak scatterings causing such fluctuation occur as follows.
Energy levels in a DWCN region are different from the corresponding levels of the leads even for a case without separation of degenerated levels because higher order terms of the intertube transfer integral exist.
This leads to similar scattering at the interfaces between leads and DWCN regions.
It is also possible that the intertube transfer causes weak backscattering.
In both cases, it is considered that an increase of the number of channels leads to an increase of backscattering because the number of final states going backward is increased.
This is a possible cause of the dependence of the conductance fluctuation on the number of channels.
\par
%
The result of nearly perfect transmission around the Fermi energy is consistent with those of previous studies \cite{Roche et al 2001a,Yoon et al 2002a}.
A study on level statistics of DWCN's showed that the mixing between states of outer and inner tubes is negligible near the Fermi energy and becomes stronger with an increase or decrease of energy \cite{Ahn et al 2003a}.
This is partly consistent with our result that reductions of the conductance occur not near the Fermi energy but in the energy region lower than that.
However, apparent asymmetric features with respect to $E\!=\!0$ cannot be seen in their results though their calculated density of states is asymmetric.
It is possible that averaging over energy windows smears the asymmetry.
\par
%
\subsection{\label{sec:level 4-2}Armchair-armchair DWCN's}
%
\begin{figure}
\scalebox{0.45}{\includegraphics{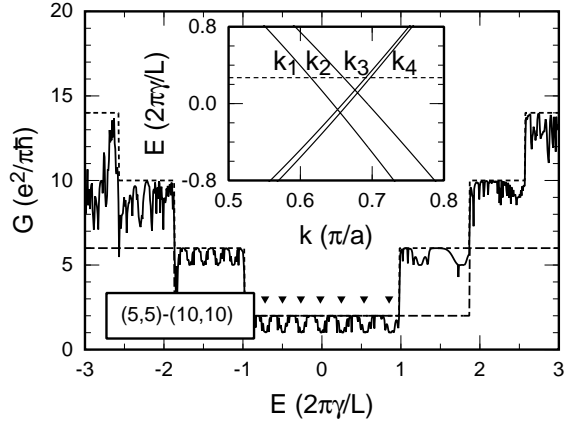}}
\caption{\label{fig:G of E (5,5)-(10,10)}
Conductance of (5,5)-(10,10) DWCN.
Inverted triangles indicate energy levels at which two beating waves become standing waves in the inner tube.
Inset shows the energy dispersion of (5,5)-(10,10) DWCN near the K point.
}
\end{figure}
%
The conductance of commensurate DWCN's depends on the structures and possibly on the circumference.
In the case of armchair-armchair DWCN's, however, there appears the characteristic common feature of conductance oscillation.
Figure~\ref{fig:G of E (5,5)-(10,10)} shows the conductance of (5,5)-(10,10) DWCN.
It can be seen that the conductance oscillates between $2e^2/\pi\hbar$ and $e^2/\pi\hbar$ for $-1\!\alt\!EL/2\pi\gamma\!\alt\!1$ and between $6e^2/\pi\hbar$ and $5e^2/\pi\hbar$ for $-2\!\alt\!EL/2\pi\gamma\!\alt\!-1$ and $1\!\alt\!EL/2\pi\gamma\!\alt\!2$.
\par
%
This is attributed to antiresonance of an incoming channel with beating standing waves in the DWCN region.
Consider the energy region with two channels.
In commensurate metal-metal DWCN's linear bands of inner tubes are degenerated to those of outer tubes in the absence of the intertube transfer.
The inset of Fig.~\ref{fig:G of E (5,5)-(10,10)} shows the energy dispersion of (5,5)-(10,10) DWCN near the K point.
In this case, level splitting of one of two linear bands is small ($k_3$ and $k_4$) but that of the other is large ($k_1$ and $k_2$).
Therefore, one incoming channel of the outer tube is not coupled to a state of the inner tube and is perfectly transmitted.
The other channel is coupled to a state of the inner tube and changes into two propagating modes with $k_1$ and $k_2$ which are extended to both the tubes.
When these two modes are localized in the DWCN region as standing waves, the conductance becomes minimum at $G\!=\!e^2/\pi\hbar$ due to antiresonance of the incoming channel with these standing waves.
\par
%
\begin{figure}
\scalebox{0.45}{\includegraphics{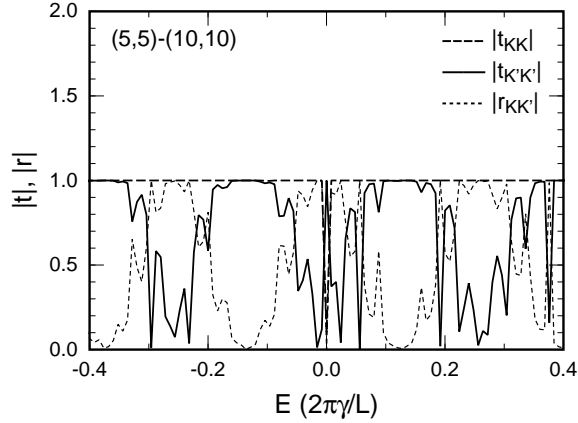}}
\caption{\label{fig:tr comm}
Energy dependence of $S$-matrix elements of (5,5)-(10,10) DWCN in a part of the energy region with two channels K and K'.
Dashed line is transmission amplitude from K to K, solid line that from K' to K', and dotted line reflection amplitude from K' to K.
}
\end{figure}
%
\begin{figure}
\scalebox{0.45}{\includegraphics{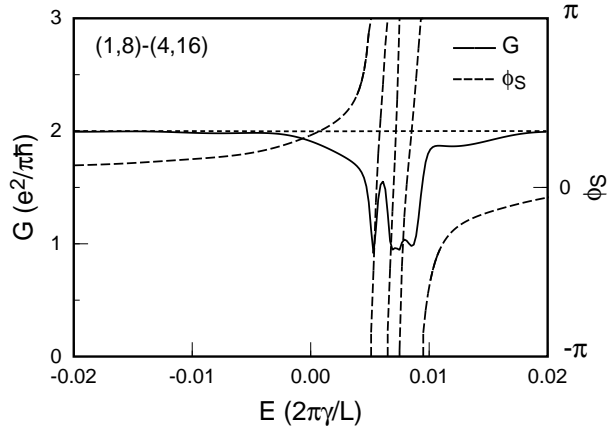}}
\caption{\label{fig:G of E near E=0} 
Conductance of (1,8)-(4,16) DWCN near $E\!=\!0$ (solid line) and the phase of the determinant of $S$ matrix (dashed line).
Dotted line is the conductance when the intertube transfer is turned off in the region within about 5\% of the length from both edges of the inner tube.
}
\end{figure}
%
The energy levels for the conductance minima are given under the condition that two beating waves with $k_1$ and $k_2$ are localized in the DWCN region, i.e., $(k_2\!-\!k_1)A/2\!=\!\pi/2\!+\!n\pi$, where $n$ is an integer.
Such energy levels are plotted by inverted triangles in Fig.~\ref{fig:G of E (5,5)-(10,10)} and are in good agreement with the conductance minima.
\par
%
Absolute values of transmission and reflection coefficients are plotted in Fig.~\ref{fig:tr comm}, in which two channels are named K and K' for convenience.
It can be seen that channel K is completely transmitted into K, while for channel K', perfect transmission into K' alternates with perfect reflection into K.
This selection rule for channels is due to the symmetry of armchair CN's \cite{Matsumura and Ando 1998a}.
This result is consistent with the above interpretation.
\par
%
\begin{figure}
\scalebox{0.45}{\includegraphics{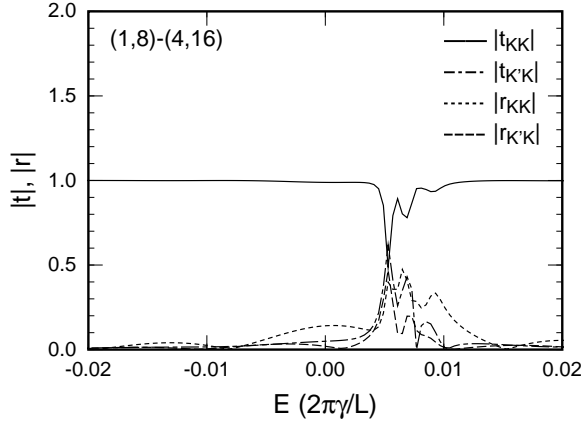}}
\caption{\label{fig:t and r}
Energy dependence of $S$-matrix elements for an incoming channel K in (1,8)-(4,16) DWCN.
Solid line is transmission amplitude to K, dot-dashed line that to K', dotted line reflection amplitude to K, and dashed line that to K'.
}
\end{figure}
%
\subsection{\label{sec:level 4-3}Antiresonance with edge states}
%
There can exist edge states at open edges of inner tubes, as mentioned in Sec.~\ref{sec:level 1}.
In this subsection, we show that antiresonance of channels in outer tubes with edge states in inner tubes leads to the conductance quantization $G\!=\!e^2/\pi\hbar$.
\par
%
Figure~\ref{fig:G of E near E=0} shows the conductance near $E\!=\!0$ as a solid line and the phase of the determinant of the $S$ matrix $\phi_S$ as a dashed line in the case of (1,8)-(4,16) DWCN.
The conductance curve has a dip structure near $EL/2\pi\gamma\!\approx\!0.007$ with minimum $G\!\approx\!e^2/\pi\hbar$.
When the intertube transfer is turned off in the region within about 5\% of the length from both edges, the dip of the conductance disappears, as shown by a dotted line.
This clearly shows that the effect arises from states localized at edges of the inner tube.
\par
%
When there are two channels interacting with an edge state, there always exists a linear combination of the two channels for which coupling with the edge state vanishes.
Although edge states are definitely degenerated in our systems, the degeneracy is lifted because of interaction with each other by way of the channels.
Looking at Fig.~\ref{fig:G of E near E=0} carefully, it can be seen that the dip consists of four resonances and the phase proceeds by $4\times 2\pi$, indicating the existence of four edge states \cite{Friedel 1953a,Moroz 1996a}.
In fact, there are four edge states in the inner (1,8) tube.
\par
%
Figure~\ref{fig:t and r} shows the transmission and reflection amplitudes for the incoming channel K.
The result shows that the conductance value $G\!\approx\!e^2/\pi\hbar$ is realized by a nontrivial combination of two channels because channel K is partly transmitted and partly reflected in a complex manner.
The result agrees well with the above interpretation.
\par
%
This antiresonance does not occur for every DWCN.
The necessary conditions are that edge states exist in inner tubes and that they have finite overlap with channels in outer tubes.
The width of dip of conductance is determined by the coupling strength between channels and edge states.
\par
%
\begin{figure}
\scalebox{0.45}{\includegraphics{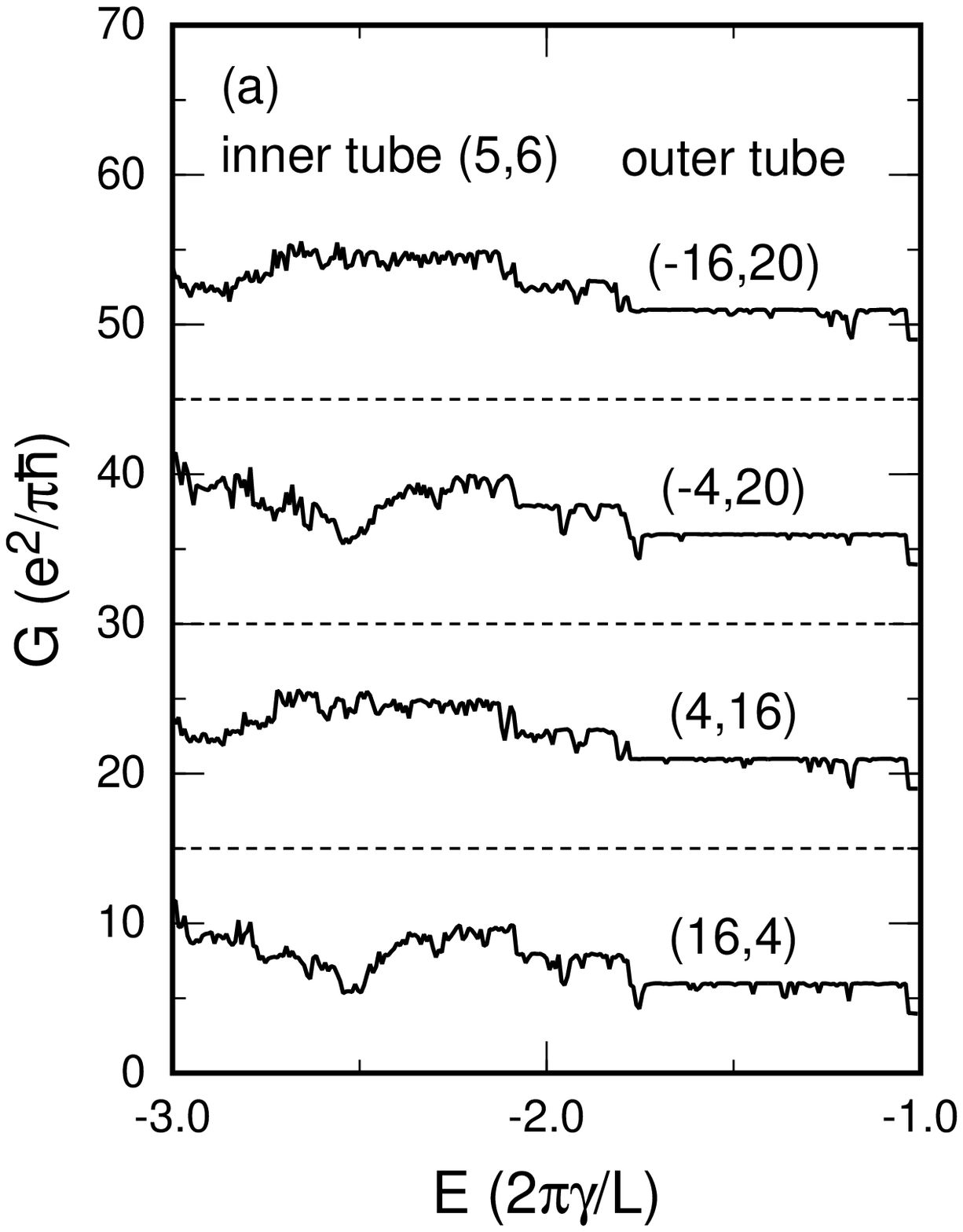}}
\scalebox{0.45}{\includegraphics{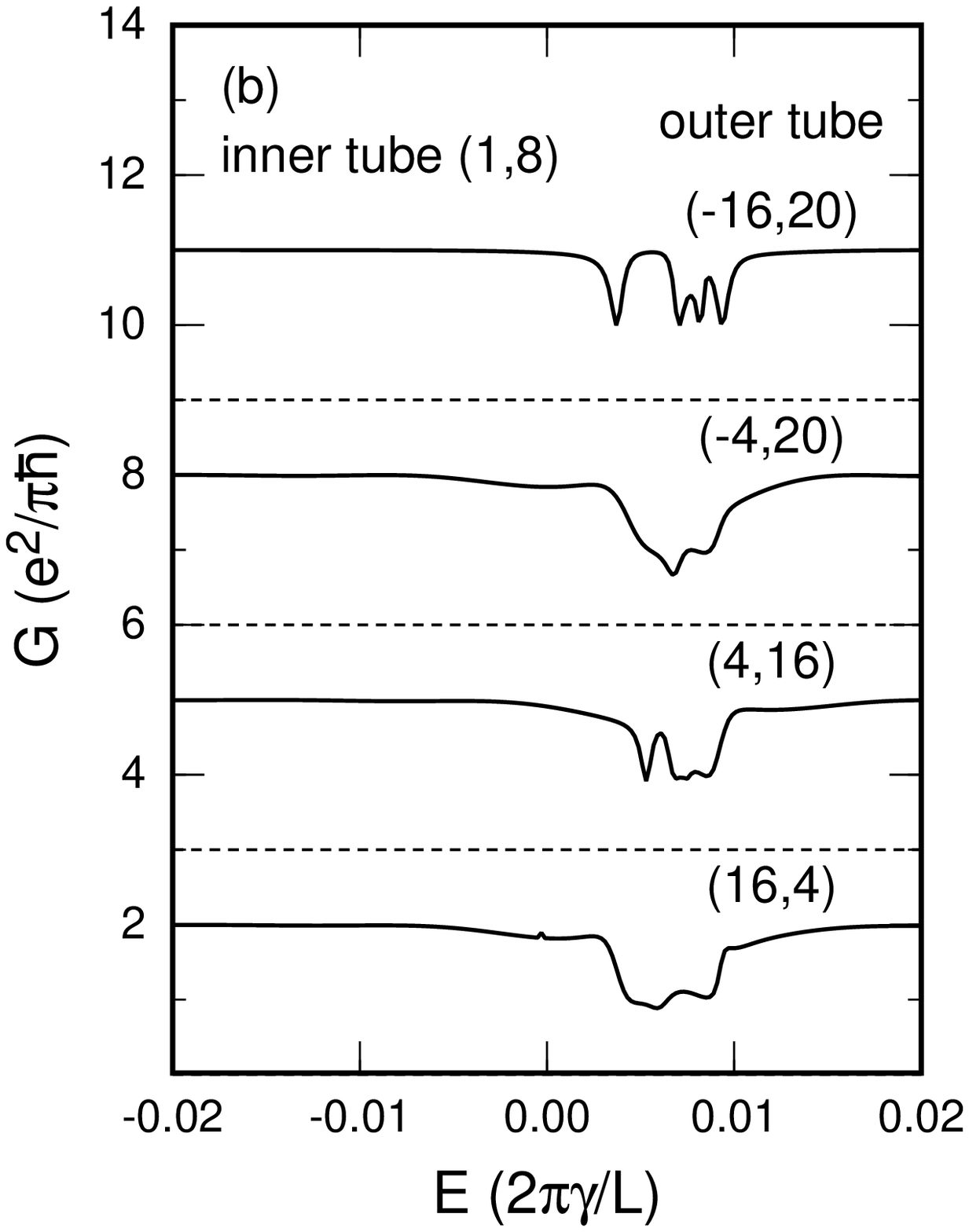}}
\caption{\label{fig:G of E for various outer CN's}
Conductance for DWCN's in which outer tubes are (16,4), (4,16), ($-$4,20), and ($-$16,20) tubes and inner tubes are the (a) (5,6) tube and (b) (1,8) tube.
In (a) the origin of the perpendicular axis is shifted by 15 for the (4,16) tube, 30 for the ($-$4,20) tube, and 45 for the ($-$16,20) tube.
In (b) it is shifted by 3 for the (4,16) tube, 6 for the ($-$4,20) tube, and 9 for the ($-$16,20) tube.
Dotted lines indicate each origin.
}
\end{figure}
%
\subsection{\label{sec:level 4-4} Cases of mirror-symmetric outer tubes}
%
In this subsection we consider three outer tubes, (16,4), ($-$4,20), and ($-$16,20) tubes, which are mirror symmetric to or the same as the (4,16) tube.
Figure~\ref{fig:G of E for various outer CN's}(a) shows the calculated conductance for an energy range in $E\!<\!0$ where the inner tube is the (5,6) tube corresponding to Fig.~\ref{fig:G of E (5,6)-(4,16) wide}.
Reductions of the conductance are seen and those for ($-$16,20) and (4,16) outer tubes or those for ($-$4,20) and (16,4) outer tubes are very similar to each other.
These results are consistent with the result that the conductance reduction is due to degenerated-level separation, because the angle between the chiral vectors for ($-$16,20) and (4,16) tubes or for ($-$4,20) and (16,4) tubes is $\pi/3$ (see Fig.~\ref{fig:structures}(c)) and the energy levels in these two cases are the same due to the six-fold rotational symmetry of energy bands of the 2D graphite sheet.
\par
%
Figure~\ref{fig:G of E for various outer CN's}(b) shows cases of antiresonance with edge states near $E\!=\!0$ where the inner tube is the (1,8) tube corresponding to Fig.~\ref{fig:G of E near E=0}.
Antiresonance occurs in all cases.
This is attributed to the fact that in the four cases, the interactions between edge states of the inner tubes and channels of the outer tubes are similar to each other.
Because among the four cases, the differences between radii of the inner and outer tubes and the momenta of the channels in the circumference and tube-axis directions are the same and only the relative lattice configurations between the inner and outer tubes are different from each other.
\par
%
Therefore, the definition of CN's using regions I to IV in Fig.~\ref{fig:structures}(c) makes the investigation of the structure dependence of the average conductance in DWCN's considerably simple.
That is to say, the average conductance is determined essentially by only DWCN's with inner tubes in region II and outer tubes in regions I and II, for example.
For antiresonance with edge states, it is sufficient to investigate DWCN's with inner and outer tubes in one region.
\par
%
\section{\label{sec:level 5}Discussion}
%
All of our results can be applied to MWCN's straightforwardly.
Therefore, we shall discuss the observation of the suppression of degenerated-level separation and antiresonance of channels with edge states in MWCN's.
So far, there does not seem to be any clear experimental evidence for those phenomena.
This is mainly because highly developed technology is required to prepare systems with negligibly small contact resistance and without disorder.
Although systems in which the conductance is quantized near the Fermi energy are necessary for the observation, there are only a few experiments on such systems \cite{Frank et al 1998a,Urbina et al 2003a}.
We believe, however, that the observation of our findings is highly possible in such systems in the following ways.
Since it is not clear how disorder in tubes affects our results, we do not consider the effects here.
That remains as an issue for future studies.
\par
%
We consider two representative experimental setups.
The first setup is that source and drain electrodes are located on or beneath tubes and a gate electrode, which controls the chemical potential of electrons in the tubes, is equipped.
The second one is that one end of the tubes is attached to the tip of a microscope and the other end is immersed in liquid metal.
\par
%
First, suppression of degenerated-level separation can be observed in the first setup as an asymmetric gate-voltage dependence of the conductance, or in both the setups as an asymmetric source-drain-voltage dependence of the conductance.
In the former case, the conductance for the positive gate-voltage region is reduced as compared to that for the negative gate-voltage region.
In the latter case, since the conductance at a finite source-drain voltage is given by the integral of the zero-bias conductance over the corresponding energy width, it also becomes asymmetric as a function of the source-drain voltage, with respect to the origin.
\par
%
However, the asymmetric source-drain-voltage dependence of the conductance was not obtained in the second setup used by Kociak {\it et al}. \cite{Kociak et al 2002a}.
It is possible that this was because the scanning range of the voltage was not sufficiently wide.
Note that although some groups reported an interesting effect of the intertube transfer of electrons at huge source-drain voltage \cite{Collins et al 2001a,Watanabe et al 2003a}, it is beyond the scope of this work because the phenomenon is related to electrical breakdown.
\par
%
Next, the observation of antiresonance with edge states is discussed.
This can be observed in both the first and second setups under the following conditions.
One is that source and drain electrodes are connected to the outer tubes.
The other is that there exist edge states in the inner tubes and the edges are located between the two electrodes.
\par
%
We must make give some comments on the edge geometry of CN's.
As is well known, caps often terminate CN's.
However, MWCN's with open edges are also realistic since MWCN's with open edges have actually been observed \cite{Iijima et al 1992a}, and it is suggested that CN's with open edges can be stable\cite{Louchev et al 2002a}.
In the case of MWCN's with caps, the following is expected.
Some previous studies revealed the existence of states localized at caps around the Fermi energy \cite{Tamura and Tsukada 1995a, Carrll et al 1997a, Yaguchi and Ando 2001a, Yaguchi and Ando 2001b}.
Therefore, similar antiresonance with such localized states may lead to the conductance quantization $G\!=\!e^2/\pi\hbar$.
There are still some issues to be clarified, for example, cap-structure dependence of eigenenergy levels of localized states and that of the coupling strength between channels and localized states.
Further studies are necessary.
\par
%
Finally note the following.
Conductance quantization was observed in a recent experiment in which the second setup with a MWCN was used and the conductance was measured as a function of the immersion depth \cite{Frank et al 1998a}.
The experiment showed that the conductance is quantized not only at $2ne^2/\pi\hbar$ with $n$ being integers but also anomalously at $e^2/\pi\hbar$ and $3e^2/\pi\hbar$.
This anomalous quantization seems to be similar to the antiresonance with edge states in our results.
However, as it is considered that the inner tubes were capped, an investigation of the possibility of antiresonance with states localized at caps is necessary.
Therefore, it cannot be concluded at present that this anomalous quantization is due to such antiresonance.
Further studies are needed to clarify this effect.
\par
%
\section{\label{sec:level 6} Summary}
%
The electronic states and the conductance of DWCN's in the absence of impurities have been systematically studied.
Scattering in the bulk is negligible and the number of channels determines the average conductance.
In general, separation of degenerated energy levels in incommensurate DWCN's is suppressed for $E\!>\!0$ and therefore, there are few effects of the intertube transfer on the conductance.
For $E\!<\!0$ large conductance reductions can occur due to the separation of degenerated energy levels.
In some DWCN's, it is possible that antiresonance with edge states in inner tubes leads to the anomalous conductance quantization $G\!=\!e^2/\pi\hbar$.
These results can be also applied to MWCN's.
\par
%
\begin{acknowledgments}
We wish to thank Professor T. Ando, Professor H. Tsunetsugu, Dr. K. Tsukagoshi, and Dr. A. Kanda for fruitful discussions and Dr. A. Kasumov and Dr. K. Tsukagoshi for critical reading of the manuscript.
Some of the numerical calculations were performed on VPP700E at the Advanced Computing Center, RIKEN.
\end{acknowledgments}
%

\end{document}